# DOES GLOBALIZATION PROMOTE OR HINDER SUSTAINABLE DEVELOPMENT? EVIDENCE FROM TÜRKİYE ON THE THREE DIMENSIONS OF GLOBALIZATION

**Emre AKUSTA***

**Abstract**

This study analyzes the impact of globalization on sustainable development in Türkiye. We used the ARDL method with annual data for the period 2000-2021. Results reveal that economic globalization promotes positively to sustainable development in the short run with a coefficient of 0.144 and in the long run with a 0.153 coefficient. Although social globalization has a negative impact with a coefficient of -0.150 in the short run, this effect turns positive with a coefficient of 0.080 in the long run. Political globalization strongly supports sustainable development with a coefficient of 0.254 in the short run and 2.634 in the long run. Finally, total globalization has a positive impact on sustainable development in the short and long run with coefficients of 0.339 and 0.196, respectively.

# KÜRESELLEŞME SÜRDÜRÜLEBİLİR KALKINMAYI DESTEKLİYOR MU YOKSA ENGELLİYOR MU? KÜRESELLEŞMENİN ÜÇ BOYUTU ÜZERİNE TÜRKİYE'DEN KANITLAR

**Öz**

Bu çalışma, küreselleşmenin Türkiye'de sürdürülebilir kalkınma üzerindeki etkisini analiz etmektedir. Çalışmada 2000-2021 dönemi yıllık verileri ile ARDL yöntemi kullanılmıştır. Sonuçlar, ekonomik küreselleşmenin kısa vadede 0,144 katsayısı ile uzun vadede ise 0,153 katsayısı ile sürdürülebilir kalkınmayı olumlu yönde desteklediğini ortaya koymaktadır. Sosyal küreselleşme kısa vadede -0.150 katsayısı ile negatif bir etkiye sahip olsa da bu etki uzun vadede 0.080 katsayısı ile pozitife dönmektedir. Politik küreselleşme kısa vadede 0,254 ve uzun vadede 2,634 katsayısı ile sürdürülebilir kalkınmayı güçlü bir şekilde desteklemektedir. Son olarak, toplam küreselleşmenin sürdürülebilir kalkınma üzerinde kısa ve uzun vadede sırasıyla 0.339 ve 0.196 katsayıları ile pozitif bir etkisi vardır.

**Anahtar kelimeler:** *Küreselleşme, Sürdürülebilir Kalkınma, ARDL Yöntemi, Zaman Serisi Analizi, Türkiye.*

*Assist. Prof. Dr., Kırklareli University, Faculty of Economics and Administrative Sciences, Department of Economics, KIRKLARELİ.
e-mail: emre.akusta@klu.edu.tr, (https://orcid.org/0000-0002-6147-5443)

# 1. INTRODUCTION

Environmental problems, depletion of finite resources and social inequality caused by human activities continue to increase worldwide. This situation has led to the questioning of the approach that development can only be measured by economic growth and sustainable development has arisen. Sustainable development aims to meet the needs of future generations while meeting the needs of the present. In this context, it has a multidimensional structure that includes economic, social and environmental dimensions (Sachs et al., 2023). Firstly, economic sustainability prioritizes long-term economic growth and the preservation of natural and social capital. Sustainability of economic activities aims to eliminate sectoral imbalances in industrial production by preserving capital and minimizing the environmental impacts of production processes. Furthermore, economic sustainability contributes to increasing social welfare by ensuring the stability and continuity of the labor market. Second, social sustainability ensures that all individuals in society have equal rights. This guarantees everyone access to basic services such as health, education, security and gender equality. Social sustainability also aims to create a fair social structure for all, emphasizing intergenerational and intra-generational equity. This includes not only today's society but also the society of future generations. Finally, environmental sustainability emphasizes the conservation and efficient use of natural resources, reducing pollution and preserving biodiversity. This dimension also includes addressing global environmental challenges by reducing the impact of environmental impact of economic operations. (Rosen, 2018; Gürlük, 2010; Atvur, 2009).

Another concept that has been much discussed and emphasized in recent times is globalization. Globalization denotes the increasing connectivity and interaction between societies and economies around the world. This process has accelerated with developments in technology, communication and international trade. Globalization has facilitated the cross-border flow of goods, services, information and capital, making world economies more interdependent. Globalization also has a multidimensional structure. Economic globalization involves the expansion of international trade, capital flows and global markets. This dimension encompasses the free movement of goods and services around the world, international investment and the activities of multinational corporations. It also includes the integration of global financial markets. Thus, it leads to a more interconnected world economy (Grinin and Korotayev, 2010; Ardalan, 2010). Cultural globalization is the spread and interaction of different cultural elements on a global scale. This process has accelerated with the development of technology and communication tools. Global media networks and digital communication platforms facilitate the cross-border transfer of cultural products and lifestyles. It also enables the rapid dissemination of ideas and information through a global communication network (Toma, 2012). Political Globalization involves the development of international political cooperation and norms. International institutions, intergovernmental agreements and transnational political movements are prominent elements of this dimension. Political globalization requires more coordinated and integrated action by the international community, especially in the search for solutions to global problems (Rifai, 2013). Technological globalization is related to the rapid spread of innovations in information and communication technologies. The Internet, social media platforms and mobile technologies are the basic building blocks of this dimension, facilitating the exchange of information and ideas worldwide. These technologies transcend the boundaries of time and space, allowing people and organizations to interact with each other (Castagna and Furia, 2010).

Globalization is also an important tool for strengthening international cooperation and multilateral relations. For example, it necessitates cooperation in tackling global challenges such as climate change and sustainable development. However, international regulations and policies need to be managed effectively to ensure a fairer distribution of the benefits of this process. Globalization brings challenges as well as opportunities. It can offer opportunities for developing countries by stimulating economic growth and accelerating the transfer of knowledge and technology. However, this process can sometimes lead to economic imbalances by increasing competitive pressure on local businesses. It can also cause problems such as cultural homogenization, environmental degradation and instability in labor markets.

Globalization has accelerated with the expansion of free market economies, increased international trade and advances in technology (Dreher et al., 2008; Jones, 2010). In addition, globalization has led to rapid changes in economic, social, technological and environmental dynamics around the world in recent years. Thus, it has encouraged the increase in international trade and the integration of national economies into the global economy (Loungani, 2005). In today's world, it is of great importance to understand the effects of these changes on sustainable development. Because sustainable development aims to increase economic and social welfare by addressing environmental challenges (Seydioğulları, 2013; Gupta and Vegelin, 2016). Moreover, assessing the compatibility of this rapid integration process with sustainable development goals is critical, especially for emerging economies such as Türkiye. On the other hand, globalization, especially with technology and capital movements, has created an interconnected and interactive economy across the world. While this process has stimulated economic growth, it has also increased environmental and social costs (Strezov et al., 2017). In this regard, developing countries need to adopt sustainable development, not just economic development.

Globalization establishes complex relationships between economic growth and social and environmental change. These relationships are shaped by a variety of factors, such as the liberalization of capital movements, trade expansion and technological progress. While these processes provide better access to international markets, they can also put pressure on local economies and increase social inequalities (Martens and Raza, 2010). Moreover, global economic integration can lead to lower environmental standards and overconsumption of natural resources. This conflicts with the basic principles of sustainable development. Therefore, environmental and social factors should also be taken into account while achieving global economic integration (Öymen, 2020). On the other hand, in the social dimension, technological changes and labor market transformations that come with globalization have led to changes in education and skill requirements. This may affect unemployment rates and social exclusion. Therefore, to achieve sustainable development goals, education and labor policies need to be adapted to these new realities. In addition, policy adjustments need to be made to ensure fairness in income distribution and to ensure that individuals benefit fairly from economic growth. Developing countries, especially Türkiye, need to consider the environmental and social costs of globalization while taking advantage of the economic opportunities it offers. Globalization's impact on sustainable development in Türkiye should be examined and the interaction between the dimensions should be carefully evaluated. Therefore, our study analyzes globalization's impact on sustainable development in Türkiye.

Our study will contribute to the literature in the following ways. (1) To the best of our knowledge, there is no empirical study that investigates the impact of globalization on sustainable development in Türkiye. This study aims to fill this gap in the literature. (2) In order to avoid the problems of multi-attribution that may arise between different dimensions of globalization, we have examined the effects of the dimensions of globalization separately. For this purpose, we have constructed 4 separate models in the study. Thus, we have more clearly analyzed the potential impact of each dimension of globalization on sustainable development. (3) We have used the most recent data set available in the study. Thus, this study provides a real-time and up-to-date perspective.

The rest of the paper presents the state of globalization and sustainable development in Türkiye, literature review, data and methodology, empirical findings and finally conclusions and policy recommendations.

## 2. THE STATE OF GLOBALIZATION AND SUSTAINABLE DEVELOPMENT IN TÜRKİYE

Globalization is causing major and transformative changes in global communities as borders become more permeable and interactions between goods, services, capital and culture increase worldwide (Çetin and Çınar, 2023). While encouraging global economic growth, this process has complex impacts on social, environmental, and economic balances. Türkiye is one of the countries experiencing globalization intensively with its economic, social and environmental impacts (Altıner et al., 2018). Sustainable development emphasizes the need to balance the opportunities for economic growth offered by globalization with environmental sustainability and social justice and offers an opportunity to reassess the challenges and opportunities that this process brings.

Türkiye's strategic position in the globalization process presents significant opportunities and challenges at both regional and global levels. Economically, while integration into global markets offers positive aspects such as expanded export capacity and increased foreign investment, it can also put pressure on domestic markets and local producers. Socially, it can affect social structures and labor markets, offering new job opportunities while simultaneously creating negative impacts on occupational safety and working conditions. Moreover, globalization, increased industrialization, and consumption pose serious challenges to environmental sustainability. With this regard, it is important to examine Türkiye's globalization and sustainable development process. Besides the economic opportunities provided by the globalization process, social justice and the preservation of environmental balance will also be important for Türkiye to realize its sustainable development goals. Therefore, in this section of our study, we examine Türkiye's globalization and sustainable development processes.

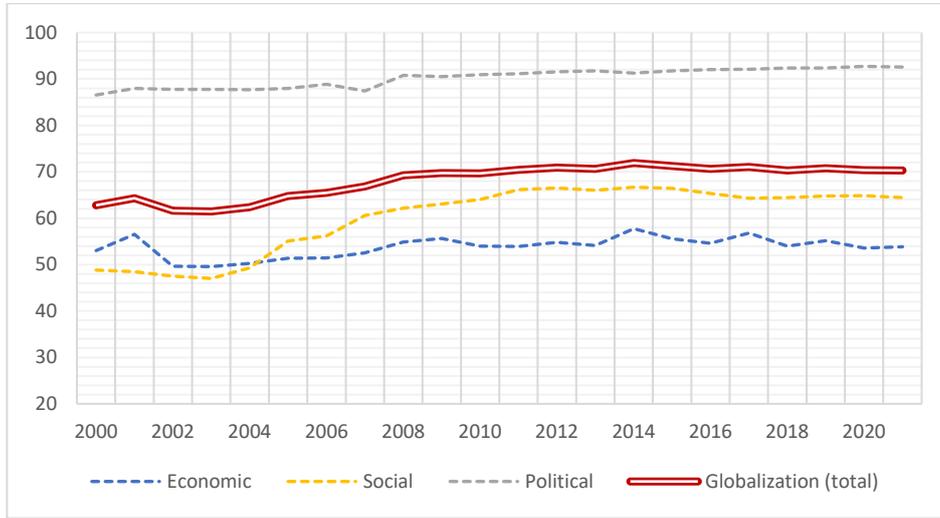

**Figure 1: Development of the Globalization Index in Türkiye**

**Source:** Constructed with Gygli (2019) data.

Figure 1 shows the changes in Türkiye's globalization index and dimensions between 2000 and 2021. The graph shows that the overall globalization index has shown relative stability and has been on an increasing trend over time. Economic globalization increased from around 50 points in the early 2000s to 55 points in 2021. This period can be attributed to factors such as Türkiye's integration into global markets and the liberalization of foreign trade and the increase in foreign direct investments. The economic globalization index showed a decline after the 2008 global financial crisis, but then showed an upward trend. Social globalization refers to cultural interactions, exchange of information and ideas, and free movement of people. The social globalization index, which was around 48 points in 2000, has steadily increased over time and reached 70 points. This increase may be a result of technological developments, innovations in communication and global cooperation in the field of education. Political globalization represents international political integration, intergovernmental cooperation and participation in international agreements. The political globalization index increased from 87 points in the early 2000s to 93 points in 2021 (Gygli, 2019). The rise in the political globalization index can be attributed to Türkiye's active role in international politics and changes in its presence in other international platforms.

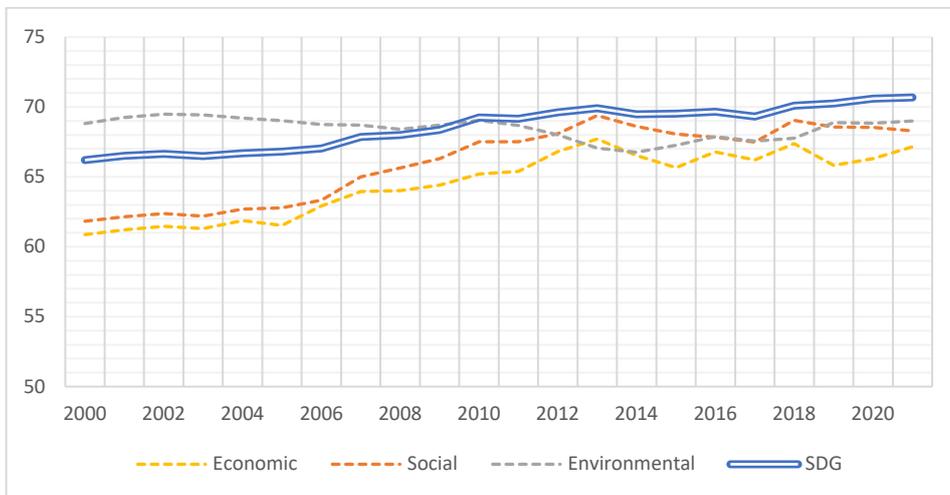

**Figure 2: Development of the SDG Index in Türkiye**
**Source:** Constructed with SDR (2023) data.

Figure 2 shows Türkiye's sustainable development goals (SDG) index and changes in its dimensions for the period 2000-2021. The SDG index shows Türkiye's progress in achieving the SDGs. The economic dimension started around 60 points in 2000 and increased significantly over time. However, it started to decline after 2013. This is related to global economic fluctuations and the wrong economic policies. The social dimension also showed a significant increase in the early 2000s but started to decline in 2013. While this increase reflects progress in areas such as education, access to health services and gender equality, the decline points to wrong policies. Moreover, the decline in the last two years points to some challenges in these areas and the impact of the pandemic on social services. The environment dimension is the most striking dimension in the graph. This is because it moves inversely to the economic and social dimensions. The environmental dimension, which was around 69 points in the early 2000s, declined until 2013 and then fluctuated continuously, rising again to 69 points in 2021. The constant fluctuations and instabilities in the environmental field reflect the complex effects of the regulations applied in these matters. These regulatory measures cover a wide range of initiatives, including the formulation and implementation of policies that promote environmental sustainability.

Moreover, SDGs index serves as a vital indicator in this context. It provides a comprehensive view of a country's progress towards achieving sustainability by providing an integrated assessment of three key dimensions: economic, social and environmental. By combining data across these dimensions, the SDG index not only highlights areas where effective measures have been implemented, but also identifies sectors where further efforts are needed. This holistic evaluation is essential for policymakers, stakeholders, and researchers engaged in designing and executing strategies for sustainable development. The total SDG index has shown a steady increase, starting with a score of 68 in 2000 and exceeding 70 in 2021 (SDR, 2023). This shows that Türkiye is generally achieving progress towards the sustainable development goals. However, fluctuations in some areas suggest that this progress has not been equal in all areas.

| Country | Turkiye |
|---|---|
| 2023 SDG Index Score | 70.8 |
| 2023 SDG Index Rank | 72 |
| SDG 1: No Poverty | ⬆ 🟡 |
| SDG 2: No Hunger | ➡ ⚪ |
| SDG 3: Good Health and Well-Being | ↗ ⚪ |
| SDG 4: Quality Education | ⬆ ⚪ |
| SDG 5: Gender Equality | ➡ 🔴 |
| SDG 6: Clean Water and Sanitation | ↗ ⚪ |
| SDG 7: Affordable and Clean Energy | ↗ 🟡 |
| SDG 8: Decent Work and Economic Growth | ➡ 🔴 |
| SDG 9: Industry, Innovation and Infrastructure | ↗ 🔴 |
| SDG 10: Reduced Inequalities | ➡ 🔴 |
| SDG 11: Sustainable Cities and Communities | ➡ ⚪ |
| SDG 12: Responsible Consumption and Production | ↗ ⚪ |
| SDG 13: Climate Action | ➡ 🔴 |
| SDG 14: Life Below Water | ↗ 🔴 |
| SDG 15: Life on Land | ➡ 🔴 |
| SDG 16: Peace, Justice and Strong Institutions | ➡ 🔴 |
| SDG 17: Partnerships for the Goals | ➡ ⚪ |

On track or maintaining achievement ⬆  Goal Achievement 🟢
Moderately Increasing ↗  Challenges remain 🟡
Stagnating ➡  Significant challenges ⚪
Decreasing ⬇  Major challenges 🔴
Insufficient data ○

**Figure 3: SDG Dashboards and Trends for Türkiye 2023**
**Source:** Constructed with SDR (2023) data.

Countries aim to reach 2030 targets in a number of critical areas, including poverty, education and gender equality, clean energy and sustainable economic growth. These comprehensive goals are important for building a more just and sustainable society on a global scale. These goals are also a roadmap for Türkiye. Figure 3 shows the indicators and trends of Türkiye's 2023 sustainable development goals. The 2023 SDG Index expresses Türkiye's status with various color codes and trend indicators. The statistics clearly show achievements as well as areas where more efforts are needed.

Türkiye's overall SDG Index score is 70.8, ranking 72nd internationally. This score shows that Türkiye has made significant progress in certain areas, but still faces some challenges that need to be overcome. Figure 3 shows moderate achievements in SDGA1 and SDG3, which focus on basic human needs. Particularly noteworthy is the rapid progress made in health care under SDG3. This progress can be seen as a reflection of investments in the health sector and efforts to increase accessible health services. However, it is clear that more sustained progress is needed in the fight against poverty (SDG1) and hunger (SDG2).

The low level of achievement in the area of quality education (SDG4) shows that more needs to be done to ensure equality of opportunity and access to quality education for all. Considering that education is a basic personal right and a key driver of a society's development, the importance of increasing investments in this area becomes even more evident. Moreover, there are major structural challenges to the environment-related SDG targets, in particular SDG13, SDG14 and SDG15. Major challenges demonstrate the urgency of these issues and the critical areas that need to be addressed. Combating climate change, sustainable water management and protecting biodiversity are crucial for Türkiye's future. Taking steps towards these goals will also contribute to sustainable development (SDR, 2023).

**3. LITERATURE REVIEW**

The origins of sustainable development date back to the mid-20th century, when the environmental and social impacts of economic activities were increasingly recognized. Since the 1960s, the negative impacts of economic growth on the environment have been discussed in public and academic communities. During this period, increasing environmental awareness raised concerns that the unlimited use of resources and intense pressure on the ecosystem were unsustainable. The concept of "sustainability" in the field of environment was first used in the early 1970s by the editors of The Ecologist journal in the UK (Gürak, 2006). In this period, the concept was mostly discussed from the perspective of environmental protection. The concept of sustainable development in today's sense was first defined in 1987 in the report entitled Our Common Future (Brundtland Report) published by the United Nations Commission on Environment and Development (WCED). This report defined sustainable development as "meeting the needs of present generations without jeopardizing the ability of future generations to meet their own needs" (WCED, 1987). This definition provided a basic framework that emphasized that economic growth should take place in harmony with the environment. Following the Brundtland Report, sustainable development has been at the center of both international policies and academic studies and has become a multidimensional field of study.

Singh (2013) states that the main objective of sustainable development is to ensure a livable world for present and future generations. To achieve this goal, wasteful consumption patterns should be abandoned, and environmental impacts should be reduced. Similarly, Capper et al. (2014) emphasize that sustainable development requires addressing social, environmental and global needs in an integrated manner. Instead of aiming only for economic growth, this approach focuses on ensuring environmental sustainability while enhancing social welfare. Sustainable development is nowadays addressed in a multidisciplinary framework. Kumar et al. (2023) and Kanivets et al. (2023) argue that sustainable development includes concrete steps such as reducing greenhouse gas emissions, waste management and promoting sustainable production and consumption patterns. This shows that sustainable development is not only a theoretical concept. At the same time, the success of this multidimensional approach depends on the cooperation of different disciplines. Góralski et al. (2020) argue that sustainable development includes an assessment of economic performance as well as an analysis of environmental impacts and savings in natural resource use. Sustainable development policies should aim to ensure efficient and renewable use of natural resources while maintaining economic growth.

Sustainability is even more important in sectors that rely on the use of natural resources, such as the agricultural sector. Schindler et al. (2015) state that sustainability impact assessments are important in reducing the risks of negative impacts, especially in the agricultural sector. Conservation of natural resources used in agricultural production ensures both environmental sustainability and food security in the long run.

The Sustainable Development Goals include 17 targets to be pursued for a globally sustainable world by 2030. These goals cover basic human needs such as ending poverty, ensuring access to quality education and clean water. They also focus on environmental challenges such as combating climate change, building sustainable cities and protecting marine and land life. The SDGs aim to promote social and environmental sustainability in balance with economic growth. Sardjono et al. (2021) and Amato (2021) explain the definition and purpose of the SDGs as a transformative set of goals designed to address challenges for people and the planet. The SDG indicators are important for tracking progress toward the achievement of the SDGs. This progress should be monitored in cooperation at all levels, private and public. Ultimately, the SDGs provide a broad framework that requires collaboration and strategic planning to solve the complex problems facing the world. This framework provides an important opportunity for transformation towards creating a sustainable future (Gigliotti et al. 2018; Gigliotti et al., 2019).

Sustainable development means realizing economic growth in a way that is compatible with environmental limits and considers social needs. In this regard, green investments play an important role. Green investments refer to investments made for the environment, renewable energy investments, investments in green areas such as increasing energy efficiency and sustainable management of natural resources. Such investments are made to reduce the carbon footprint, combat climate change and contribute to the preservation of ecological balance. Green investments also provide economic and social benefits such as stimulating economic growth, creating new jobs and increasing social welfare. Green investments constitute one of the cornerstones of sustainable development with the goal of leaving a more livable world for future generations. Mantaeva et al. (2021) argue that renewable energy technologies are essential for economic growth and environmental protection. Liu et al. (2023) analyze the impacts of green investments on China's energy consumption structure, while Peng et al. (2023) detail the positive impacts on regional green ecological levels. Xiong and Dai (2023) and Wan and Sheng (2022) provide evidence on the positive impacts of green investments on sustainable development, showing that these investments provide both environmental and economic benefits.

Another important issue is globalization. Globalization has affected all regions of the world through factors such as the expansion of international trade, increased exchange of information and technology, and increased interaction between cultures. This process directly affects sustainable development goals by creating economic and social transformations that cross borders. The interactive relationship between globalization and sustainable development can be both negative and positive. Moreover, examining the effects of globalization on sustainable development is important for future policies and ensuring fair and balanced development on a global scale. Therefore, the issue of sustainable development and globalization has been examined in many ways in the literature (see, for example, Tekbaş, 2019; Pekar, 2020; Koyuncu and Karabulut, 2021; Bilgili et al., 2022; Polat and Ergün, 2023, Ojaghlou and Tercan, 2024). It is important to comprehend the dimensions of globalization and its impact on sustainable development. In especially, being aware of the opportunities provided by globalization and distributing these opportunities fairly is the key to achieving a sustainable future on a global scale. In this context, a review of the existing literature reveals the various impacts of globalization and its relationship with sustainable development. It also provides an understanding of both the opportunities globalization offers and the challenges it poses. Therefore, we reviewed studies that investigate the impacts of globalization on environmental, economic and social dimensions and their implications for sustainable development.

Globalization has various impacts on the environment. Figge et al. (2017) and Lee and Min (2014) argue that globalization increases environmental pressures by increasing the ecological footprint. In particular, the globalization index has a negative impact on environmental sustainability by leading to the intensification of economic activities and increased consumption. Moreover, Postolache et al. (2019) argue that economic globalization affects the environment through structural and compositional effects. Studies in Türkiye, Central and Eastern European countries, transition economies and BRICS countries provide valuable contributions to the understanding of the relationship between economic growth, energy consumption, globalization and environmental indicators. Destek and Özsoy (2015) analyzed the validity of the EKC hypothesis for Türkiye. The study finds that energy consumption and economic growth increase environmental degradation, while economic globalization reduces CO2 emissions. Similarly, Destek (2020) examined the environmental impacts of different dimensions of globalization in Central and Eastern European countries. The study emphasized the role of economic and social globalization in increasing carbon emissions while political globalization in reducing environmental pollution. Both studies confirm the validity of the EKC hypothesis. Tekbaş (2021), in his study on transition economies, revealed the positive impact of economic growth and economic globalization on CO2 emissions. This relationship is supported by the increasing effect of energy consumption. Moreover, the study found bidirectional causality between economic growth and globalization and CO2 emissions. These findings emphasize that free market economic policies should focus on environmentally friendly technologies and energy efficiency. Finally, Pata et al. (2024) examined the impacts of income, globalization and technological progress on ecological footprint in BRICS countries. The study showed that economic growth increases the ecological footprint, while globalization has a decreasing impact on many components. However, technological progress does not have a significant impact. The study concluded that economic development and environmental sustainability need to be balanced. Overall, while these studies address the environmental impacts of globalization, economic growth and energy consumption through different geographical and methodological approaches, they show that the EKC hypothesis is confirmed in many contexts. The extent of the environmental impacts of globalization is found to vary depending on the diversity of policies and economic structures. This suggests that regional and country-specific strategies should be developed for environmental sustainability.

Moreover, globalization also has economic impacts. The economic dimensions of globalization have positive effects such as increased output and employment as well as negative effects such as increased income inequalities. Kandil et al. (2017), in their study comparing China and India, found that globalization has a positive impact on economic growth in India and a negative impact in China. In addition, unidirectional causality was found from economic growth to globalization in India and bidirectional causality in China. These findings highlight the differences in the economic structures and globalization processes of the two countries. Altıner et al. (2018), in their study on 10 emerging market economies, find that the impact of social globalization on economic growth varies across countries. For example, while social globalization has a positive impact on growth in Argentina, Indonesia, Mexico and Russia, it has a negative impact in Poland. The study did not find a direct causality from social globalization to economic growth, but emphasized that economic growth affects social globalization.

Hassan et al. (2019), analyzing Pakistan, found that globalization has a negative impact on economic growth in the long run. This finding is supported by the existence of a unidirectional causality relationship from economic growth to globalization. This result suggests that Pakistan has not fully achieved the expected economic benefits of globalization. Çelik and Ünsür (2020), in a panel data analysis of 88 countries, found a reciprocal causality relationship between economic growth and economic, social and technological globalization. However, the existence of a unidirectional causality relationship from economic growth to general and political globalization is one of the important findings of the study. This shows that different dimensions of globalization have a complex relationship with growth. Tekbaş (2022) analyzed the relationship between economic growth and economic, social and political globalization in BRICS-T countries and found that all dimensions of globalization have a positive impact on growth. Moreover, a unidirectional causality relationship was found between economic and social globalization and growth, while a bidirectional causality relationship was found between capital accumulation and political globalization. These results indicate that globalization policies in BRICS-T countries contribute to economic growth. Das (2022) theoretically and empirically examined the effects of globalization on income distribution and economic sustainability. Moreover, Behera and Sahoo (2023), in their study on 133 countries, found that globalization contributes more to human development in high-income countries. Gasimli et al. (2022) found a positive impact of globalization on sustainable development in some regions. Cervantes et al. (2020) investigated the dynamics of globalization in countries with different income levels. They also analyzed the effects of globalization on the economy. The study found a relationship between globalization and health expenditures and public expenditures. However, this relationship varies for different income groups. Moreover, globalization affects social sustainability through its effects on income distribution and job opportunities.

Sertyesilisik (2022) emphasizes that globalization has a major influence on social sustainability. Ojeyinka and Osinubi (2022) analyzed the impacts of social globalization on sustainable development in Africa. The research results indicate that social globalization has a negative impact on development goals and therefore, appropriate policies and strategies need to be identified to deal with these impacts. Moreover, the effects of globalization differ according to geographical regions. Behera and Sahoo (2023) emphasize the importance of international policies and programs for low-income countries to fully benefit from the advantages of globalization.

Globalization can affect various aspects of sustainable development in both positive and negative ways. An analysis of these impacts can help us better understand the various effects of globalization and develop policies that are appropriate to the opportunities and challenges that this process brings. These impacts vary across countries and time periods. Therefore, country-specific conditions and needs should be considered when examining the impacts of globalization.

## 4. DATA AND METHODOLOGY

### 4.1. Model Specification and Data

The empirical research of this study investigates the impact of globalization on sustainable development in Türkiye. We used annual data for the period 2000-2021. We determined the period of the study based on data availability. In order to avoid multicollinearity problems that may arise between different dimensions of globalization, we examined the effects of the dimensions of globalization separately. For this purpose, we constructed 4 separate models in the study. Thus, we analyzed the impact of globalization and each of its dimensions on sustainable development more clearly. The models created for the empirical research are shown in Equations 1-4.

**Model 1:** $SDI_{(M-D)} = \alpha_0 + \beta_1 ECON_t + \beta_2 GDP_t + \beta_3 OPEN_t + \beta_4 ACCOU_t + \beta_5 CONSMP_t + \mu_t$ (1)

**Model 2:** $SDI_{(M-D)} = \alpha_0 + \beta_1 SOCI_t + \beta_2 GDP_t + \beta_3 OPEN_t + \beta_4 ACCOU_t + \beta_5 CONSMP_t + \mu_t$ (2)

**Model 3:** $SDI_{(M-D)} = \alpha_0 + \beta_1 POLIT_t + \beta_2 GDP_t + \beta_3 OPEN_t + \beta_4 ACCOU_t + \beta_5 CONSMP_t + \mu_t$ (3)

**Model 4:** $SDI_{(M-D)} = \alpha_0 + \beta_1 GLOB_t + \beta_2 GDP_t + \beta_3 OPEN_t + \beta_4 ACCOU_t + \beta_5 CONSMP_t + \mu_t$ (4)

In Equations 1-4, $\alpha_0$ is the constant term of the models and represents the intercept terms. $\beta_1$, $\beta_2$, $\beta_3$, …, $\beta_8$ are the slope coefficients written next to each variable. is the error term. The symbol t below the variables represents the time period. The Sustainable Development Index (SDI), which plays a central role in our model, has been determined as the dependent variable. This index is an indicator developed by Sachs et al. (2023) that measures the sustainable development performance of countries. The main explanatory variables of the model include globalization and its various dimensions. Globalization is represented by the KOF globalization index developed by Dreher (2006). The other explanatory variables considered in the study are selected from the literature review and are derived from studies such as Onaran and Boesch (2014), Liu and Meissner (2015), Grechyna (2020), Oliveira et al. (2021), Maji (2022), Khan et al. (2023), Shi et al. (2023). These variables are important for understanding different aspects of globalization and their interactions. All indicators used in our study and their descriptive statistics are presented in Table 1.

**Table 1: Descriptive Statistics (2000-2021)**

| Variable | Notation | Description | Mean | Median | Min. | Max. | Std. Dev. | Source |
|---|---|---|---|---|---|---|---|---|
| Sustainable Development | SDI | Index | 68.55 | 69.21 | 66.20 | 70.68 | 1.49 | SD |
| Economic Globalization | ECON | Index | 53.76 | 53.97 | 49.57 | 57.75 | 2.18 | KOF |
| Social Globalization | SOCI | Index | 60.09 | 64.18 | 47.02 | 66.67 | 7.08 | KOF |
| Political Globalization | POLI | Index | 90.27 | 91.03 | 86.55 | 92.72 | 2.02 | KOF |
| Globalization | GLOB | Index | 68.04 | 70.00 | 61.46 | 71.90 | 3.49 | KOF |
| GDP per capita | GDP | cons. 2015 US$ | 9308 | 8808 | 5994 | 13450 | 2203 | WB |
| Trade | OPEN | % of GDP | 52.18 | 50.21 | 42.35 | 71.08 | 6.74 | WB |
| Current account balance | ACCOU | % of GDP | -3.50 | -3.83 | -8.87 | 1.86 | 2.45 | WB |
| Final consumption expenditure | CONSMP | % of GDP | 74.93 | 75.40 | 68.19 | 78.87 | 2.52 | WB |

Note: (1) SD, KOF, and WB indicate Sustainable Development Report-Country Profiles, Swiss Economic Institute-KOF Index, and World Bank-World Development Indicators, respectively. (2) The logarithm of all variables has been taken.

### 4.2. Unit Root Analysis

In time series analysis, it is critical to determine whether the series are stationary in order to make reliable estimates. Previous studies have shown that economic time series often have non-stationary properties (Johansen and Juselius, 1990). Autocorrelation should be a property that should decrease as lags increase. However, in non-stationary series, this autocorrelation does not converge to zero but remains far from zero. This is an indication that the series is non-stationary. This situation in non-stationary series leads to a problem called "spurious regression" in econometric analysis. Spurious regression is a misleading situation where independent variables appear to have a significant effect on the dependent variable, but in fact this relationship is completely random. This can cause estimates to be misleading and unreliable. This may lead to biased and inconsistent forecasts (Granger and Newbold, 1974). Therefore, it is necessary to check whether time series are stationary or not.

Unit root tests are used to determine whether a time series is stationary. If the series is non-stationary, it is seen that the probabilistic (stochastic) process changes over time and the series is non-stationary. In a stationary time series, the difference between two consecutive values is independent of time. To ensure stationarity, if the series becomes stationary when a certain number of differences are taken, the series is called time-integrated and denoted as I(d). For example, a series integrated at the zeroth order is stationary and denoted as I(0), while a series differenced at the first order is denoted as I(I) (Kennedy, 2006: 356).

In this study, Augmented Dickey-Fuller (ADF) and Phillips-Perron (PP) unit root tests are applied to determine whether the series are stationary. While the ADF test tests stationarity by considering possible autocorrelation in the series, the PP test tests stationarity by taking into account autocorrelation and variance in the error terms of the series.

The ADF unit root test is evaluated through a regression model that analyzes the relationship between the series itself and its previous values. The model of the ADF test is expressed as in Equation 5 (Kutlar, 2007):

$$\Delta y_t = \alpha + \beta_t + \gamma y_{t-1} + \sum_{i=1}^{p} \varphi_i \Delta y_{t-1} + \varepsilon_t \qquad (5)$$

In Equation 5, $\Delta y_t$ is the first difference of the series, $y_{t-1}$ is the previous value of the series, $\alpha$ is the constant term, $\beta_t$ is the trend term and $\epsilon_t$ is the error term. The null hypothesis of the ADF test is that the series has a unit root.

The PP unit root test, similar to the ADF test, tests for the presence of a unit root in the series. However, instead of modeling autocorrelation and heteroskedasticity in the series, it uses a method that corrects the variance of the series. The PP unit root test is expressed as in Equation 6:

$$y_t = \rho y_{t-1} + \varepsilon_t \tag{6}$$

In Equation 6, the coefficient indicates the correlation of the series with previous values and the error term considers autocorrelation and heteroskedasticity. If the test statistics calculated in both tests are greater than the critical values, the series are stationary. In other words, if the test statistics are greater than the critical value, the series do not contain unit root. These tests are necessary for the correct application of the ARDL model.

**4.3. ARDL Cointegration Test**

The Autoregressive Distributed Lag (ARDL) method is a frequently preferred method in econometric analyses because it can be used in samples with a limited number of observations and variables can have different degrees of integration. This method, developed by Pesaran et al. (2001), allows the variables to be analyzed without the requirement that they are integrated to the same degree. The ARDL method has the capacity to estimate both short-run and long-run relationships. In other words, variables can be included in the analysis as stationary at the level, at the first difference or at the first difference of another variable while one variable is at the level. In this respect, while the ARDL method allows variables to be I(0) or I(I), it is not suitable for I(2) and more highly integrated series (Pata et al., 2016).

The ARDL method can be performed without the need for any unit root test, but unit root tests are required to prove that the variables are non-stationary in their second differences. This is because appropriate table critical values are not available when the variables are stationary in the second difference. In conclusion, the ARDL cointegration test is an effective method to test for the existence of long-run relationships between variables and offers advantages over the Johansen-Juselius (1990) and Engle-Granger (1987) cointegration tests as it allows flexibility in the degree of integration of variables.

The ARDL cointegration test is constructed to include past values of the dependent variable and both past and current values of the explanatory variables. The ARDL model is as in Equation 7 (Pesaran et al., 2001):

$$Y_t = a_0 + \sum_{i=1}^{p} a_i Y_{t-1} + \sum_{j=0}^{q} \beta_{j1} X_{1,t-j} + \cdots + \sum_{j=0}^{q} \beta_{jk} X_{k,t-j} + \epsilon_t \tag{7}$$

The ARDL cointegration test is used to determine the existence of long-run relationships between series. The main purpose of this test is to see whether the series move together. The calculated test statistic is compared with the upper bounds table. If the calculated F statistic exceeds the upper bound value in the table, the null hypothesis is rejected. This means that the variables are cointegrated (Narayan and Smith, 2005).

**4.4. ARDL Long-Run Estimation**

Estimation of the long-run coefficients follows the determination of the cointegration relationship between the data. In this study, the long-run relationships between the series are estimated using the ARDL model and considering the specified lag lengths. The long-run coefficients in the cointegration equation of the ARDL model are used for long-run estimations. The model established for this purpose is shown in Equation 8:

$$Y_t = \frac{\beta_{10} X_{1,t} + \beta_{20} X_{2,t} + \cdots + \beta_{k0} X_{k,t}}{1 - \sum_{i=1}^{p} \alpha_i} + \epsilon_t \tag{8}$$

The optimal lag lengths in the models are determined using the Akaike Information Criterion (AIC) and the Schwarz Information Criterion (SIC) proposed by Pesaran et al. (2001). When making long-run forecasts, it is assumed that the model is cointegrated and balanced relationships are correctly estimated. Moreover, it is considered that the lag length should not exceed two for annual data. Various diagnostic tests are applied to assess the accuracy of the model. These tests examine the model for normal distribution, autocorrelation and constant variance. In addition, it was also assessed that the functional form of the model was properly established and that the estimated parameters were reliable.

## 5. EMPIRICAL FINDINGS

In the initial phase of the analysis, unit root tests were applied to the variables utilized in the study, with the objective of determining their stationarity. The results of these tests are presented in Tables 2, which include the models that include only the constant term and those that include both the constant and the trend term, respectively.

**Table 2: Unit Root Test Results**

| Variable | | ADF unit root test | | PP unit root test | |
|---|---|---|---|---|---|
| | | t-statistic (level) | t-statistic (first difference) | t-statistic (level) | t-statistic (first difference) |
| SDI | Constant | -1.792 | -2.439** | -0.748 | -5.149*** |
| ECON | | -2.531 | -6.801*** | -2.480 | -6.801*** |
| SOCI | | -1.919 | -2.071** | -1.498 | -2.700*** |
| POLI | | -0.944 | -8.251*** | -1.595 | -8.860*** |
| GLOB | | -6.786** | -2.120*** | -1.377 | -4.157*** |
| GDP | | 0.054 | -4.408*** | 0.065 | -4.473*** |
| OPEN | | -0.730 | -5.357*** | -0.458 | -5.714*** |
| ACCOU | | -3.008* | -6.079*** | -3.002* | -8.375*** |
| CONSMP | | -0.050 | -4.221*** | 0.615 | -4.102*** |
| SDI | Constant and Trend | -3.046 | -3.693** | -1.755 | -5.018*** |
| ECON | | -3.073 | -6.634*** | -3.073 | -6.634*** |
| SOCI | | -1.070 | -3.041** | -0.736 | -3.101*** |
| POLI | | -3.357* | -8.088*** | -3.451* | -8.732*** |
| GLOB | | -1.468 | -4.088*** | -0.943 | -4.088** |
| GDP | | -3.501* | -4.315*** | -3.120 | -4.412*** |
| OPEN | | -2.297 | -4.887*** | -2.386 | -12.935*** |
| ACCOU | | -2.902 | -4.990*** | -2.894 | -15.673*** |
| CONSMP | | -1.555 | -4.270** | -1.400 | -4.497*** |

Note: The superscripts ***, **, and * denote the significance at a 1%, 5%, and 10% level, respectively.

Table 2 illustrates that, when the constant term is included in the unit root test results, the majority of variables are found to be stationary in their first differences. Additionally, the GLOB and ACCOU variables were found to be stationary at the level. However, the SDI, ECON, SOCI, POLI, GDP, OPEN, and CONSMP variables were found to exhibit a unit root at the level but to become stationary at the first difference. In addition, Table 2 also presents the results of the models with constant and trend. In particular, the POLI and GDP variables are stationary at both the level and first difference. The remaining variables are not stationary at the level but become stationary at the first difference. In summary, the unit root test results indicate that the majority of the variables are I(I), suggesting greater flexibility in cointegration testing methods such as the ARDL bounds test.

**Table 3: ARDL Bounds Test Results for Cointegration**

| Model | Optimal lag length | F-statistics | Critical values %5 | | Critical values %1 | |
|---|---|---|---|---|---|---|
| | | | $I(0)$ | $I(I)$ | $I(0)$ | $I(I)$ |
| Model 1:F(SDI | ECON, GDP, OPEN, ACCOU, CONSMP) | (2, 2, 0, 2, 1, 2) | 7.545*** | 2.39 | 3.38 | 3.06 | 4.15 |
| Model 2:F(SDI | SOCI, GDP, OPEN, ACCOU, CONSMP) | (2, 2, 2, 1, 2, 2) | 10.108*** | 2.39 | 3.38 | 3.06 | 4.15 |
| Model 3:F(SDI | POLIT, GDP, OPEN, ACCOU, CONSMP) | (1, 2, 2, 2, 2, 2) | 4.235*** | 2.39 | 3.38 | 3.06 | 4.15 |
| Model 4:F(SDI | GLOB, GDP, OPEN, ACCOU, CONSMP) | (2, 2, 2, 2, 1, 2) | 9.142*** | 2.39 | 3.38 | 3.06 | 4.15 |

Note: The superscripts ***, **, and * denote the significance at a 1%, 5%, and 10% level, respectively.

Table 3 presents the results of the ARDL bounds test. The results include the optimal lag lengths for the model, F-statistic values, and critical values at both the 5% and 1% significance levels. The F-statistics of the models employed in this study exhibit a considerable range, spanning from 4.235 to 10.108. All of these values exceed both of the critical values. Consequently, it can be concluded that there is a statistically significant and long-run relationship between the variables in the model. These findings indicate the existence of a robust and statistically significant cointegration between the dependent variable and the explanatory variables. Consequently, in the final stage of the analysis, we estimate the short-run and long-run coefficients. The results of these estimations are presented in Table 4.

Before estimating the short-run and long-run coefficients, the models constructed in the study were subjected to a series of specification tests. A serial correlation LM test, histogram-normality test, heteroskedasticity test, and Ramsey RESET test were conducted to determine whether the model was free from autocorrelation, variable variance, specification errors, and normality distribution. Furthermore, the CUSUM and CUSUM of squares tests were employed to identify structural breaks. The results of these tests indicate that there are no specification errors in the model and that the F-statistics and slope coefficients of the estimated models are reliable. The test results are presented in Table 5. The plots of the CUSUM and CUSUMSQ tests are presented in Figure 4. These results demonstrate that the model is robust and reliable.

**Table 4: Short-Run and Long-Run Estimation**

| Dependent variable: SDI$_{(M-D)}$ | Short-run coefficients | | | |
|---|---|---|---|---|
| Regressors | Model 1 | Model 2 | Model 3 | Model 4 |
| ECON | 0.144*** | | | |
| SOCI | | -0.150*** | | |
| POLI | | | 0.254** | |
| GLOB | | | | 0.339*** |
| GDP | 0.251*** | 0.359*** | 0.098** | 0.201*** |
| OPEN | 0.062*** | 0.029*** | 0.084*** | 0.038** |
| ACCOU | -0.002* | -0.011*** | -0.002 | -0.004** |
| CONSMP | 0.344*** | 0.390*** | 0.535*** | 0.250*** |
| ECT(-1) | -0.372*** | -0.145*** | -0.358*** | -0.438*** |
| Dependent variable: SDI$_{(M-D)}$ | Long-run coefficients | | | |
| Regressors | Model 1 | Model 2 | Model 3 | Model 4 |
| ECON | 0.153*** | | | |
| SOCI | | 0.080*** | | |
| POLI | | | 2.634** | |
| GLOB | | | | 0.196*** |
| GDP | 0.069*** | 0.057*** | 0.284 | 0.047** |
| OPEN | 0.109*** | 0.081*** | 0.223 | 0.078*** |
| ACCOU | -0.004*** | -0.009*** | -0.032** | -0.004** |
| CONSMP | 0.309*** | 0.332*** | -2.086 | 0.239** |
| C | 0.554*** | 0.700*** | 2.1279** | 0.705*** |
| Diagnostic tests | P value | P value | P value | P value |
| $\chi^2$ (Serial correlation) | 0.17 | 0.20 | 0.14 | 0.47 |
| $\chi^2$ (Heteroskedasticity) | 0.71 | 0.64 | 0.42 | 0.74 |
| $\chi^2$ (Normality) | 0.76 | 0.65 | 0.88 | 0.69 |
| $\chi^2$ (Functional form) | 0.43 | 0.18 | 0.24 | 0.32 |
| CUSUM | Stable | Stable | Stable | Stable |
| CUSUMSQ | Stable | Stable | Stable | Stable |

Note: The superscripts ***, **, and * denote the significance at a 1%, 5%, and 10% level, respectively.

**MODEL I**

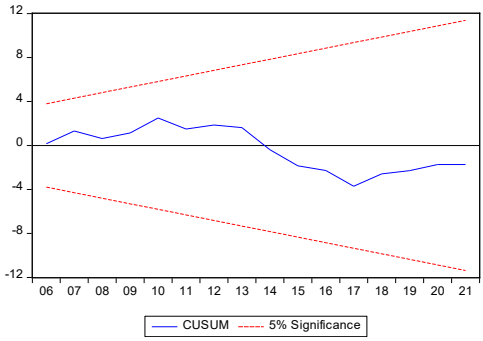 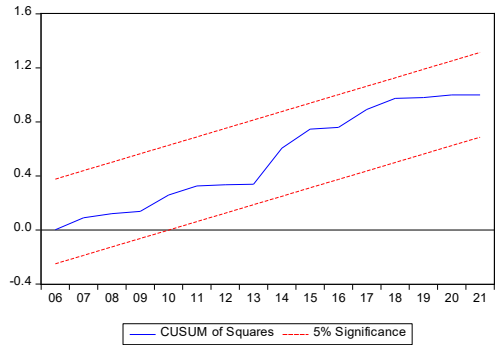

**MODEL II**

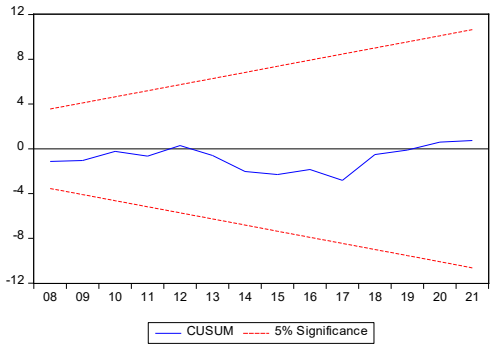 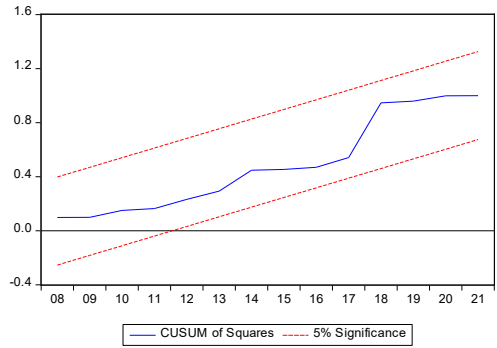

**MODEL III**

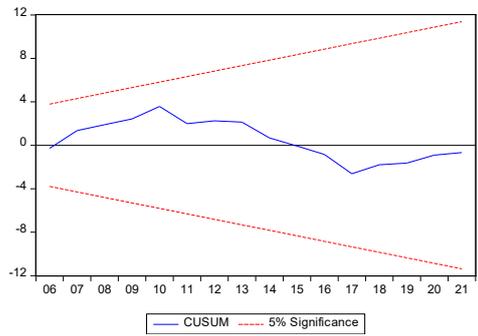 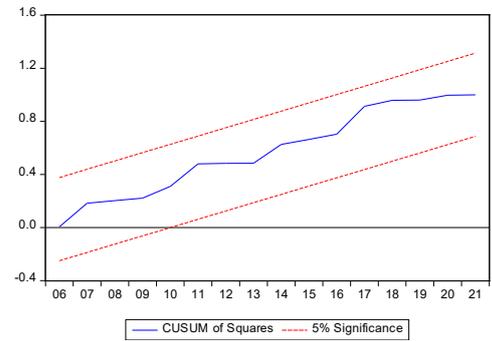

**MODEL IV**

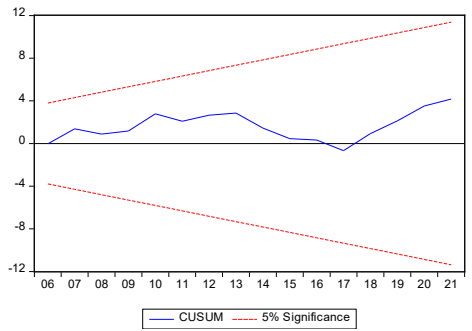 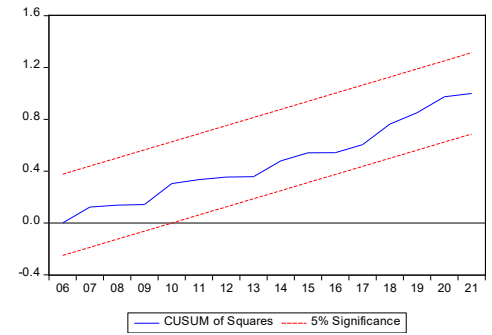

**Figure 4. Plots of CUSUMSQ and CUSUM Statistics**

Table 4 shows the impact of globalization and its different dimensions on sustainable development. The results show that economic globalization has a positive impact on sustainable development in the short term, with a coefficient of 0.144. This influence increases over time, which is reflected in a coefficient of 0.153 in the long run. The short-term benefits of economic globalization, such as enhanced free trade, increased investment flows, and broader market access, become more substantial over time. Conversely, social globalization is initially detrimental to sustainable development, as indicated by a coefficient of -0.150 in the short run. In the long run, however, this effect turns positive with a coefficient of 0.080, suggesting that social integration and cultural exchange, despite initial challenges, ultimately support sustainable development. Political globalization shows a positive effect from the beginning, with a coefficient of 0.254 in the short run, which increases significantly to 2.634 in the long run. This implies that political cooperation and integration have profound and long-lasting positive effects on sustainable development.

The results of Model 4 indicate that in the short run, GDP per capita has a strong and positive impact on sustainable development, as evidenced by a coefficient of 0.201. In the long run, the impact diminishes but remains significant at 0.047, indicating that while economic growth has a rapid initial response, it tends to stabilize over time. In particular, investments in short-term economic growth have a pronounced effect on sustainable development. In addition, foreign trade has a positive impact on sustainable development in the short run, with a coefficient of 0.038, and this impact becomes stronger in the long run, reaching a coefficient of 0.078, indicating that foreign trade not only promotes sustainable development quickly, but also provides greater benefits over time. Conversely, the current account balance has a negative impact on sustainable development in both time periods, with a coefficient of -0.004, suggesting that current account deficits are consistently detrimental to sustainable development, underscoring the importance of reforms and policies aimed at improving the current account balance. In addition, final consumption expenditure contributes positively and significantly to sustainable development in the short run, with a coefficient of 0.250, and this positive impact persists in the long run, indicating that increased consumer confidence and spending can both accelerate economic growth in the short run and enhance sustainable development in the long run.

Total globalization has a comprehensive impact on sustainable development by including economic, social and political dimensions together. The results of Model 4 show that in the short run, total globalization has a very strong and positive impact on sustainable development with a coefficient of 0.339. The results show that the combined effect of different aspects of globalization promotes sustainable development through economic growth, social integration and political cooperation. At the same time, international cooperation, investment and cultural interactions can have positive consequences for economic and social welfare as well as environmental sustainability. In the long run, the coefficient of 0.196 on the impact of aggregate globalization is again positive but lower than in the short run. This decline reflects the time-varying effects of various aspects of globalization and perhaps some challenges. In the long run, factors such as political and economic instability, environmental costs or socioeconomic imbalances may make it difficult to sustain initially positive economic and social benefits. For example, free trade may hurt local industries in the long run, social integration may lead to cultural conflicts, or political cooperation may destabilize some regions. Therefore, the declining long-run impact of total globalization may be an indicator of complex dynamics and rising costs. This makes it imperative for policymakers to carefully consider all aspects of globalization and develop long-term strategies while preserving short-term gains. Such an approach is critical to maximizing the positive impacts of globalization on sustainable development and minimizing potential negative consequences.

The findings of this study are both similar and different from the literature on the impacts of globalization on sustainable development. The positive impact of economic globalization on sustainable development in the short and long run is consistent with Gasimli et al. (2022) and Behera and Sahoo (2023). These studies emphasized the supportive impacts of globalization on economic growth and human development. However, studies such as Destek (2020) and Tekbaş (2021) emphasize the negative environmental impacts of economic globalization. This difference suggests that the role of economic globalization in sustainable development should be considered multidimensionally. The negative impact of social globalization in the short term and positive impact in the long term supports the findings of Ojeyinka and Osinubi (2022) on the initial negative impacts of social globalization. However, a distinctive finding of our study is that this impact turns positive in the long run. This suggests that social integration processes can contribute to sustainable development over time. The positive and strong impact of political globalization is consistent with Gasimli et al. (2022) and Support (2020). These studies also emphasize that political cooperation and integration make deep contributions to sustainable development in the long run. However, it should be noted that policies may differ regionally.

The findings related to the impacts of the explanatory variables on sustainable development are broadly consistent with the literature. While GDP per capita has a strong positive impact in the short run, this impact diminishes and stabilizes in the long run. This finding is in line with Behera and Sahoo (2023) and Pata et al. (2024), who find that economic growth promotes sustainable development in the short run, but its impact diminishes in the long run due to environmental costs. The positive impact of trade in the short run becomes stronger in the long run. This result is consistent with Gasimli et al. (2022) who find that trade promotes economic growth and sustainable development. However, Tekbaş (2021) emphasized that trade may have environmental costs in the long run. This suggests that trade policies should be designed with a focus on sustainability. The current account deficit has a negative impact on sustainable development in both the short and long run. This is in line with the general view in the literature that problems such as economic instability and resource scarcity make sustainable development difficult. Behera and Sahoo (2023) support this finding by emphasizing that this impact is more pronounced in low-income countries. In conclusion, the impacts of these variables are generally consistent with the literature. In this respect, more comprehensive and balanced policies need to be developed to minimize environmental and social costs.

## 6. CONCLUSION

In this study, the impact of globalization on sustainable development in Türkiye from 2000 to 2021 was investigated. The ARDL approach was used to assess the presence of cointegration among variables and to determine their coefficients. This method provided an opportunity to examine the impact of globalization on sustainable development, which is the central focus of our research. In addition, to mitigate potential problems of multicollinearity arising from the different dimensions of globalization, we analyzed each dimension separately. To achieve this goal, four different models were developed to examine the specific impact of each dimension of globalization on sustainable development.

The results of our research show that economic globalization provides positive and significant benefits on sustainable development. In particular, integration into global markets and increased foreign capital flows contribute to sustainable development by increasing the growth potential of local economies. These interactions enable the transformation of economic structures through technology transfer and the adoption of innovative business practices. However, the impact of social globalization on sustainable development is more complex. Our research shows that social globalization can have negative impacts on sustainable development in the short run, but in the long run, these impacts turn positive. It shows that increased cultural interactions and the expansion of social networks can turn short run negatives into positives. Initial challenges stem from processes of cross-cultural adaptation and the integration of local communities with global values. Over time, however, these processes can increase societies' openness to innovation and strengthen their awareness of environmental and social responsibility.

Political globalization has a consistent and positive impact on sustainable development in both the short and long run. It facilitates this through mechanisms such as international cooperation, political integration, higher environmental standards, the implementation of sustainable energy policies, and unified efforts to address global issues. In addition, political globalization promotes sustainability goals through international regulations and agreements. Our research also examines the impact of economic indicators on sustainable development. We find that in addition to the positive short- and long-term effects of GDP per capita and foreign trade, the current account balance and final consumption expenditure also significantly enhance sustainable development. Specifically, improvements in the current account balance support sustainable development by promoting economic stability and increasing resilience to external financial disturbances. In addition, growth in final consumption expenditures stimulates domestic production and employment, which promotes economic growth and increases overall social welfare.

The results of the study demonstrate that globalization has a positive impact on sustainable development. Apparently, strategic policies need to be developed to reduce potential risks while maximizing the opportunities offered by globalization. Several policies have been formulated to achieve this goal. (1) To enhance the positive impacts of economic globalization on sustainable development, liberalized trade should be promoted, investment flows should be increased and market access opportunities should be expanded. Trade policies and investment incentives should be revised accordingly. (2) To strengthen the long-term positive impact of social globalization, policies that promote social integration should be developed. There is also a need for long-term planning of population movements and migration. Alongside this planning, programs to increase cultural interactions and policies that embrace social diversity need to be developed. In this way, social conflicts can be reduced and society can become more inclusive. (3) International political cooperation should be promoted to enhance the impact of political globalization on sustainable development. Reinforcing diplomatic relations and promoting regional integration are important in this respect. (4) Economic policies should be reconsidered to manage the effects of current account balance and consumption expenditures on sustainable development. In particular, measures should be taken to ensure that foreign trade policies contribute to the current account balance. (5) Long-term strategies should be developed to ensure the continuation of the positive effects of globalization on sustainable development. These strategies should include a comprehensive approach for future generations, balancing economic, social and environmental factors. These policy recommendations can contribute to achieving Türkiye's sustainable development goals and maximize the positive effects of globalization while minimizing potential risks.

Although this study provides important findings, it has some limitations. First, this study has analyzed sustainable development as a whole. Different dimensions of sustainable development are not emphasized. Future studies can expand on this aspect and assess it from a broader perspective. Second, future research could expand the sample to include both developed and developing countries. Thus, it can compare the results between these two groups. It can be investigated whether the effects of globalization on sustainable development differ according to the level of economic development. Finally, larger data sets and different analysis techniques can be used to test the reliability of the findings of this study. Future research could improve the study by following these directions.

**Disclosure Statements (Beyan ve Açıklamalar)**

1. The authors of this article confirm that their work complies with the principles of research and publication ethics (Bu çalışmanın yazarları, araştırma ve yayın etiği ilkelerine uyduklarını kabul etmektedirler).

2. No potential conflict of interest was reported by the authors (Yazarlar tarafından herhangi bir çıkar çatışması beyan edilmemiştir).

3. This article was screened for potential plagiarism using a plagiarism screening program (Bu çalışma, intihal tarama programı kullanılarak intihal taramasından geçirilmiştir).